\documentclass[pra,aps,showpacs,preprint]{revtex4}
\begin{document}

\title{Theory for quantum state of photon pairs generated from
spontaneous parametric down conversion nonlinear process}
\author{Ruo Peng WANG}
\email{rpwang@cis.pku.edu.cn}
\author{Hui Rong ZHANG}
\affiliation{ Physics Department and State Key Laboratory for
mesoscopic Physics, Peking University, Beijing 100871, P.R.China }

\date{\today}

\begin{abstract}

We present a theory for the quantum state of photon pairs generated
from spontaneous parametric down conversion nonlinear process in
which the influence of the final sizes of nonlinear optical crystals
on eigen optical modes is explicitly taken into consideration. We
find that these photon pairs are not in entangled quantum states.
Polarization correlations between the signal beam and the idler beam
are explained. We also show that the two photons generated from SPDC
are not spatially separated, therefore the polarization correlation
between the signal and idler beams is not an evidence for quantum
non-locality.

\end{abstract}


\maketitle


\section{Introduction}

The entangled quantum state is the core of Eistein-Podolsky-Rosen
(EPR) paradox \cite{epr}, and forms a base for many possible
application of quantum information. Photon pairs generated from
spontaneous parametric down conversion (SPDC) nonlinear process are
generally considered as entangled photon pairs. They are widely used
in experiments that require entangled photon pairs as light sources.
By applying a quantum transition theory, Shi and co-workers showed
that photon pairs generated from SPDC nonlinear process are quantum
entangled \cite{shih}. But however, in their theory, the finite
sizes of nonlinear optical crystals were not taken into
consideration. According to their simplification, the eigen modes of
optical field are plan waves, even in the presence of optical
crystals. But as pointed out by N.Bloembergen \cite{bloe}, the
correct eigen modes of optical field, in this case, are linear
combinations of plan waves determined by boundary conditions at the
surfaces of optical crystals. Because the quantum state of photons
generated from SPDC nonlinear process depend explicitly on the eigen
optical modes, the correctness of Shi and co-workers' conclusion on
the quantum state of photon pairs generated from SPDC is a question.

In this paper, we present a quantum theory for the quantum state of
photon pairs generated from SPDC, in which the effect of the finite
size of optical crystals is explicitly taken into consideration. We
find that these photon pairs are not in entangled quantum states,
and the correlation between the polarizations of photons is not an
evidence for quantum non-locality.

A hamiltonian that describes SPDC process is introduced in
Sec.\ref{sec:ham}, and the eigen optical modes are analyzed in
Sec.\ref{sec:mode}. An explicit expression for the quantum state of
photon pairs generated from SPDC is established in
Sec.\ref{sec:state}. This expression for the quantum state is used
in Sec.\ref{sec:cor} for analyzing the polarization correlation
between the signal and idler beams.

\section{The Hamiltonian for spontaneous parametric down conversion process}
\label{sec:ham}

Let's consider an optical crystal with the second order optical
nonlinearity. The effective Hamiltonian for the optical parametric
process is given by \cite{shih}:
\begin{eqnarray}{\label {ham}}
    H_1 = \varepsilon_0 \int_V \sum_{i,j,h =1}^3 \chi_{lmn}
    E_{pi}^{(+)}E_{sj}^{(-)} E_{sh}^{(-)} d^3 \vec r \;+\; h. c.
\end{eqnarray}
where   $\varepsilon_0$ is the dielectric permittivity of the
vacuum, $V$ is the volume of the optical crystal, $\chi$ is the
second order nonlinear electric susceptibility tensor, $\vec
E_p^{(+)}$ is the positive frequency part of the pump optical field,
$\vec E_s^{(-)}$ is the negative frequency part of the generated
optical field, and $h. c.$ means the Hermite conjugate.

We expand optical fields $\vec E_p^{(+)}$ and $\vec E_s^{(-)}$ into
linear combinations of eigen optical modes:
\begin{eqnarray}{\label{exp}}
    \vec E_{p}^{(+)}=\sum_{\omega_p,l} b_l(\omega_p) \vec u_{l}(\omega_p,\vec r)
    \exp (-i\omega_p t)
    \; \mbox{ and }\;
    \vec E_{s}^{(-)}=\sum_{\omega_s,l} b^\dag_l(\omega_s)
    \vec u^*_{l}(\omega_s,\vec r) \exp (i\omega_s t),
\end{eqnarray}
where $b_l(\omega)$ is the annihilation operator for a photon with
the circle frequency $\omega$ in the eigen mode $\vec
u_{l}(\omega,\vec r)$. The eigen mode $\vec u_{l}(\omega,\vec r)$
satisfied the following orthonormality conditions:
\begin{eqnarray}\label{ort1}
    \int \sum_{j,k =1}^3 \varepsilon_{jk}
    u_{lj} (\omega,\vec r)u_{mk}^{*}(\omega,\vec r)d^3 \vec r
    = \frac{1}{2}\hbar \omega \delta_{lm},
\end{eqnarray}
and
\begin{eqnarray}\label{ort2}
    \int \sum_{j,k =1}^3 \varepsilon_{jh}
    u_{lj} (\omega_1,\vec r)u_{mh}^{*}(\omega_2,\vec r)d^3 \vec r
    = 0,\;\mbox{ if }\;\omega_1 \neq \omega_2 .
\end{eqnarray}
By applying relations (\ref{exp}),(\ref{ort1}) and (\ref{ort2}), we
obtain the following expression for the total Hamiltonian:
\begin{eqnarray}{\label {ham}}
    H = \sum_{\omega}\sum_{l}\hbar \omega b^\dag_l(\omega)b_l(\omega)+
    H_1,
\end{eqnarray}
with
\begin{eqnarray}{\label {ham1}}
    H_1 = \sum_{l,m,m^\prime} M(l,m,m^\prime)
    b_l(\omega_p)b^\dag_m(\omega_s)
    b^\dag_{m^\prime}(\omega_s^\prime)
    \exp [-i(\omega_p - \omega_s - \omega_s^\prime ) t]
    \;+\; h. c.,
\end{eqnarray}
where
\begin{eqnarray}\label{mat}
    M(l,m,m^\prime) = 2 \varepsilon_0 \int_V \sum_{i,j,h =1}^3 \chi_{ijh}
    u_{li} (\omega_p,\vec r)u_{mj}^{*}(\omega_s,\vec r)
    u_{m^\prime h}^{*}(\omega_s^\prime, \vec r) d^3 \vec r .
\end{eqnarray}

According to the quantum transition theory, SPDC may occurs only
$\omega_p = \omega_s + \omega_s^\prime $ and $M(l,m,m^\prime) \neq
0$. To calculate the matrix element $M(l,m,m^\prime)$, one needs
explicit expression of eigen optical mods. One may observe that due
to the presence of a optical crystal of a finite size, plan waves
are no longer eigen modes of the optical field.

\section{Eigen optical modes}
\label{sec:mode}

We consider eigen optical modes in presence of a $L_x \times L_y
\times L_z$ uniaxial optical crystal, with the optical axis oriented
in the direction $\vec e_a = (\sin \theta, 0, \cos \theta)$.

According the exprssion (\ref{mat}), one needs only the expression
for eigen optical modes in the optical crystal. Within the optical
crystal, the eigen optical modes are linear combinations of plan
waves that are reflected into each other at the crystal's surface.
One may observe that the constant of normalization is not important
in determining the quantum state of photon pairs generated from
SPDC.

In real applications, the optical axis is so oriented that the
integration (\ref{mat}) is significantly different from zero only
for optical modes containing plan waves component with $|k_z| \gg
|k_x|, |k_y|$. These plan waves are totally reflected at surfaces
with $x=\pm L_x/2$ and at surfaces with $y=\pm L_y/2$, but weakly
reflected at surfaces with $z=\pm L_z/2$. We will neglect
reflections at $z=\pm L_z/2$. We also neglect the coupling between
{\em o}-beams and {\em e}-beams due to reflections at $x=\pm L_x/2$
and $y=\pm L_y/2$, because the principal plans are either nearly
parallel or nearly perpendicular to the plans of incidence in these
cases.

By using these approximations, we may separate eigen optical modes
into {\em o}-modes and {\em e}-modes. We have the following
expression for the $m-$th {\em o}-modes, within the crystal:
\begin{eqnarray}\label{omd}
    \vec u_o(k_{mx},k_{my};\vec r) = N_o(m)\sum_{l=1}^{4} \vec e_o(\vec k_{ml})
    \exp (i \vec k_{ml} \cdot \vec r + i \phi_{l}),
\end{eqnarray}
where $\phi_{l}$ are phase factors determined by boundary conditions
at $x=\pm L_x/2$ and $y=\pm L_y/2$, $N_o(m)$ a normalization
constant, $k_{mx}>0,k_{my}>0$,
\begin{eqnarray}
    \vec k_{m1} =(k_{mx},k_{my},k_{oz}),\;
    \vec k_{m2} =(k_{mx},-k_{my},k_{oz}),\;
    \vec k_{m3} =(-k_{mx},k_{my},k_{oz}),\;
    \vec k_{m4} =(-k_{mx},-k_{my},k_{oz}),\;
\end{eqnarray}
with $k_{oz} = \sqrt{n_o^2 k_0^2 -k_{mx}^2 -k_{my}^2}$, where $k_0 =
\omega/c$ and $n_o$ is the refractive index of {\em o}-beams, and
$\vec e_o(\vec k_{ml})$ a vector of unity that is perpendicular to
$\vec k_{ml}$ and $\vec e_a$. Similarly, for the $m-$th {\em
e}-modes, we have
\begin{eqnarray}\label{emd}
    \vec u_e(k_{mx},k_{my};\vec r) = N_e(m)\sum_{l=1}^{4} \vec e_e(\vec k_{ml})
    \exp (i \vec k_{ml} \cdot \vec r + i \phi_{l}),
\end{eqnarray}
where $N_e(m)$ is a normalization constant,
\begin{eqnarray}
    \vec k_{m1} =(k_{mx},k_{my},k_{ez}),\;
    \vec k_{m2} =(k_{mx},-k_{my},k_{ez}),\;
    \vec k_{m3} =(-k^\prime_{ex},k_{my},k_{ez}),\;
    \vec k_{m4} =(-k^\prime_{ex},-k_{my},k_{ez}),\;
\end{eqnarray}
with $k_{ez}$ is the z-component of the wave vector of an {\em
e}-beam with x and y-component of the wave vector given by $k_{mx}$
and $k_{my}$, and $-k_{ex}^\prime$ the x-component of the wave
vector of an {\em e}-beam with z and y-component of the wave vector
given by $k_{ez}$ and $k_{my}$
($k_{mx}>0,k_{ex}^\prime>0,k_{my}>0$). $\vec e_e(\vec k_{ml})$ is a
vector of unity that defines the polarization of an {\em e}-beam
with the wave vector $\vec k_{ml}$. Due to the symmetry of the
crystal, eigen optical modes can be classified into even modes with
\begin{equation}\label{meven}
    \exp (i \phi_1) = \exp (i \phi_2)
    \mbox{ and } \exp (i \phi_3) = \exp (i \phi_4)
\end{equation}
and odd modes with
\begin{equation}\label{modd}
    \exp (i \phi_1) = - \exp (i \phi_2)
    \mbox{ and } \exp (i \phi_3) = - \exp (i \phi_4)
\end{equation}
In calculations for quantum correlation between two photons
generated from SPDC, we need also the expression for optical field
outside the crystal. For optical modes with $k_x = 0$ , we have,
\begin{eqnarray}\label{omdo}
    \vec u_o(0,k_y;\vec r) &\propto& \left(\vec e_y +
    \vec e_x \frac{k_y}{n_o k_0}\cot\theta
    -\vec e_z \frac{k_y}{k_0} \right)
    \exp (i k_z z + i k_y y) \nonumber \\
    &&+
    \left(\vec e_y - \vec e_x \frac{k_y}{n_o k_0}\cot \theta
    + \vec e_z \frac{k_y}{k_0}\right)
    \exp (i k_z z - i k_y y + i \phi),
\end{eqnarray}
and
\begin{eqnarray}\label{emdo}
    \vec u_e(0,k_y;\vec r)&\propto&  \left(\vec e_x
    - \vec e_y \frac{k_y}{n_o k_0}\cot\theta \right)
    \exp (i k_z z + i k_y y) \nonumber \\
    &&+
    \left(\vec e_x + \vec e_y \frac{k_y}{n_o k_0}\cot \theta \right)
    \exp (i k_z z - i k_y y + i \phi),
\end{eqnarray}
with $k_z = \sqrt{k_0^2 - k_y^2}$.

\section{The quantum state of photon pairs}
\label{sec:state}

Suppose that the pump photon is in the quantum state
$|\psi_p\rangle$, and the generated photon pair is in the quantum
state $|\psi\rangle$. According to quantum transition theory, we
have the following relation between $|\psi\rangle$ and
$|\psi_p\rangle$:
\begin{equation}\label{stt1}
    |\psi\rangle \propto \sum_{l,m,m^\prime} \sum_{\omega_s}
    M(l,m,m^\prime)b_l(\omega_p)b^\dag_m(\omega_s)
    b^\dag_{m^\prime}(\omega_p-\omega_s)|\psi_p\rangle .
\end{equation}
In the case that pump beam is a plan wave propagating in the
direction $\vec e_z$, we have
\begin{eqnarray}
    \sum_{l} b_l(\omega_p) \vec u_{l}(\omega_p,\vec r)|\psi_p\rangle
    =\vec e_p \exp (i k_p z)|0\rangle
\end{eqnarray}
According to expressions (\ref{mat}) and (\ref{stt1}), we obtain
then
\begin{eqnarray}\label{stt2}
    |\psi\rangle &\propto& \sum_{m,m^\prime} \sum_{\omega_s}
    \int_V \sum_{i,j,h =1}^3 \chi_{ijh}
    e_{pi}u_{mj}^{*}(\omega_s,\vec r)
    u_{m^\prime h}^{*}(\omega_s^\prime, \vec r)
    \exp (i k_p z) d^3 \vec r
    \nonumber \\
    &&b^\dag_m(\omega_s)b^\dag_{m^\prime}(\omega_p-\omega_s)|0\rangle .
\end{eqnarray}
There are two types of phase match conditions \cite{phm}: type-I,
where the integral in (\ref{stt2}) is significantly different from
zero only if both modes are either $o$-modes or $e$-modes, and
type-II, where one of these two modes is an $o$-mode, and another is
an $e$-mode. We will consider only the type-II phase match in this
paper. The case of type-I phase match can be treated in a similar
way.

In the case of type-II phase match, the expression (\ref{stt2}) can
be written as
\begin{eqnarray}\label{stt3}
    |\psi\rangle &\propto& \sum_{m_o,m_e} \sum_{\omega_s}
    \sum_{l,l^\prime=1}^4
    \int_V \sum_{i,j,h =1}^3 \chi_{ijh}
    e_{pi}e_{oj}(\vec k_{m_o l})e_{eh}(\vec k_{m_e l^\prime })
    \nonumber \\
    && \exp [i( k_p z - \vec k_{m_ol}\cdot \vec r - \vec k_{m_e l^\prime}
    \cdot \vec r -\phi_l - \phi_{l^\prime})] d^3 \vec r \;
    b^\dag_{m_o}(\omega_s)b^\dag_{m_e}(\omega_p-\omega_s)|0\rangle
    .
\end{eqnarray}
If the optical crystal is large enough, then the integral in the
above expression is significantly different from zero only for mode
pairs satisfying the condition
\begin{equation}\label{pmc}
    \vec k_{m_o 1} + \vec k_{m_e 4} = k_p \vec e_z
    \mbox{  or  }
    \vec k_{m_o 2} + \vec k_{m_e 3} = k_p \vec e_z .
\end{equation}
In these cases, we have
\begin{eqnarray}\label{stt4}
    |\psi\rangle &\propto& \sum_{m_o} \sum_{\omega_s}
    c(m_o,m_e)
    b^\dag_{m_o}(\omega_s)b^\dag_{m_e}(\omega_p-\omega_s)|0\rangle
    ,
\end{eqnarray}
where
\begin{equation}
    c(m_o,m_e)=\left\{\begin{array}{cc}
                    1 & \mbox{if both of }m_o\mbox{ and }m_e
                    \mbox{ are even or odd} \\
                    0 & \mbox{other cases}
                  \end{array}\right. ,
\end{equation}
and $m_e$ is determined by the condition (\ref{pmc}).

According to the expressions (\ref{omdo}) and (\ref{emdo}), one may
observe that the optical fields of an $o$-mode and an $e$-mode may
perfectly overlap outside the optical crystal if
$k_{m_ox}=k_{m_ex}=0$, $k_{m_oy}=k_{m_ey}=k_{yd}$ and $\omega_p = 2
\omega_s$. If Eq. (\ref{pmc}) is satisfied by these modes, then this
kind of photon pairs can be generated by SPDC. By using the
expression (\ref{stt4}), we find the quantum state for such a photon
pair as
\begin{eqnarray}\label{stt5}
    |\psi\rangle =
    b^\dag_o(0,k_{yd})b^\dag_{e}(0,k_{yd})|0\rangle
    .
\end{eqnarray}
We used $(k_{mx},k_{my})$ as the label of the mode in the above
expression.

The optical field of these photon pairs is an overlap of $
b_o(0,k_{yd})\vec u_o(0,k_{yd};\vec r)$ and $b_{e}(0,k_{yd})\vec
u_e(0,k_{yd};\vec r) $. We have, outside the optical crystal
\begin{equation}\label{fld}
    \vec E^{(+)} (\vec r) \propto \vec {\cal E}_s (\vec r)
    + \vec {\cal E}_i (\vec r),
\end{equation}
where
\begin{equation}\label{fld1}
    \vec {\cal E}_s (\vec r) =\left [ b_o (0,k_{yd})
    \left(\vec e_{ys} + \vec e_x \frac{k_{yd}\cot \theta }{n_o k_0}
    \right)
    + b_e(0,k_{yd})
    \left(\vec e_x - \vec e_{ys} \frac{k_{yd}\cot \theta }{n_o k_0}\right)
    \right ]
    \exp (i \vec k_s \cdot \vec r)
\end{equation}
and
\begin{equation}\label{fld2}
    \vec {\cal E}_i (\vec r) =\left [ b_o (0,k_{yd})
    \left(\vec e_{yi} - \vec e_x \frac{k_{yd}\cot \theta }{n_o k_0}
    \right)
    + b_e(0,k_{yd})
    \left(\vec e_x + \vec e_{yi} \frac{k_{yd}\cot \theta }{n_o k_0}\right)
    \right ]
    \exp (i \vec k_i \cdot \vec r + i \phi),
\end{equation}
with $\vec e_{ys} = \vec e_{y} - \vec e_{z}k_{yd}/k_0$, $\vec e_{yi}
= \vec e_{y} + \vec e_{z}k_{yd}/k_0$, $\vec k_s = (0, k_{yd},
\sqrt{k_0^2 - k_{yd}^2})$ and $\vec k_i = (0, - k_{yd}, \sqrt{k_0^2
- k_{yd}^2})$ .

In the most of real experiments, pump beams are Gaussian beams. We
have in these cases
\begin{eqnarray}
    \sum_{l} b_l(\omega_p) \vec u_{l}(\omega_p,\vec r)|\psi_p\rangle
    =\vec e_p \int \exp \left(-\frac{k_x^2 + k_y^2}{2 \sigma^2}
    \right)
    \exp [i (\sqrt{k_p^2 - k_x^2 - k_y^2} z + k_x x + k_y y)]
    dk_x dk_y|0\rangle ,
\end{eqnarray}
and the quantum state for photon pairs with $\omega_s =1/2 \omega_p$
is given by
\begin{eqnarray}\label{stt6}
    |\psi\rangle &\propto& \sum_{m_o,m_e}
    \left[\exp \left(-\frac{(k_{m_o1x}+k_{m_e4x})^2 + (k_{m_o1y}+k_{m_e4y})^2}
    {2 \sigma^2}\right)\right.
    \nonumber \\ &&
    +\left.\exp \left(-\frac{(k_{m_o2x}+k_{m_e3x})^2 + (k_{m_o2y}+k_{m_e3y})^2}
    {2 \sigma^2}
    \right)\right]
    \nonumber \\
    &&
    \delta(k_p -k_{m_o1z} - k_{m_e1z})
    b^\dag_o(m_o)b^\dag_{e}(m_e)|0\rangle
    .
\end{eqnarray}
$k_{mx}$ and $k_{my}$ have discretely values , with $\Delta k_x \sim
1/L_x$ and $\Delta k_y \sim 1/L_y$. Because $L_x$ and $L_y$ are much
greater than the wavelength, so we may replace the summation over
$m$ by integration over $k_x$ and $k_y$. Let
\begin{equation}
    \pm k_{m_ox}\rightarrow k_{ox},\;
    \pm k_{m_oy}\rightarrow k_{oy} +k_{yd},\;
    k_{m_ex},-k^\prime_{ex}\rightarrow k_{ex},\;
    \pm k_{m_ey}\rightarrow k_{ey} +k_{yd},
\end{equation}
we may rewrite expression (\ref{stt6}) as
\begin{eqnarray}\label{stt7}
    |\psi\rangle &\propto& \int d k_{ox} \int d k_{ex}
    \int d k_{oy} \int d k_{ey}
    \exp \left(-\frac{(k_{ox}-k_{ex})^2 + (k_{oy}-k_{ey})^2}
    {2 \sigma^2}\right)
    \nonumber \\
    &&
    \delta(a_{ox}k_{ox} + a_{ex}k_{ex} + a_{oy}k_{oy} + a_{ey}k_{ey})
    b^\dag_o(k_{ox},k_{oy})b^\dag_{e}(k_{ex},k_{ey})|0\rangle
    ,
\end{eqnarray}
where
\begin{equation}
    a_{o,ex} = \left.\frac{\partial k_{o,ez}(k_{o,ex},k_{yd})}{\partial k_{o,ex}}
    \right|_{k_{o,ex}=0},\;
    a_{o,ey} = \left.\frac{\partial k_{o,ez}(0,k_{o,ey}+ k_{yd})}{\partial
    k_{o,ey}}
    \right|_{k_{o,ey}=k_{yd}} .
\end{equation}

The optical field of the photon pairs outside of the optical crystal
in the case of a Gaussian pump beam is given by
\begin{equation}\label{fld3}
    \vec E^{(+)} (\vec r) \propto \vec {\cal E}_s (\vec r)
    + \vec {\cal E}_i (\vec r),
\end{equation}
where
\begin{eqnarray}\label{fld4}
    \vec {\cal E}_s (\vec r) &=&
    \int d k_x \int d k_y
    \left [ b_o (k_x,k_y + k_{yd})
    \left(\vec e_{ys} + \vec e_x \frac{(k_{yd}+k_y-k_x)\cot \theta }{n_o k_0}
    \right) + b_e(k_x,k_y+k_{yd})\right.
    \nonumber \\ &&
    \left.
    \times \left(\vec e_x - \vec e_{ys} \frac{(k_{yd}+k_y-k_x)\cot \theta }{n_o k_0}\right)
    \right ]
    \exp (i \vec k_s \cdot \vec r)
    \exp [i(k_x x + k_y y)]
\end{eqnarray}
and
\begin{eqnarray}\label{fld5}
    \vec {\cal E}_i (\vec r) &=&
    \int d k_x \int d k_y
    \left [ b_o (k_x,k_y + k_{yd})
    \left(\vec e_{yi} - \vec e_x \frac{(k_{yd}+k_y-k_x)\cot \theta }{n_o k_0}
    \right) + b_e(k_x,k_y+k_{yd})\right.
    \nonumber \\ &&
    \left.
    \times \left(\vec e_x + \vec e_{yi} \frac{(k_{yd}+k_y-k_x)\cot \theta }{n_o k_0}\right)
    \right ]
    \exp (i \vec k_i \cdot \vec r + \phi)
    \exp [i(k_x x + k_y y)]
\end{eqnarray}

\section{Polarization correlation}
\label{sec:cor}

Having the expression for the quantum state and optical field
outside the optical crystal for photon pairs generated from SPDC, we
can analyze now the polarization correlation between the signal and
idler beams. For simplicity, we consider the case of a plan wave
pump beam.

According to quantum transition theory, the probability $P(\alpha,
\beta)$ of finding simultaneously a photon in the signal beam (the
optical beam the wave vector $\vec k_s$) with a polarization in the
direction $\vec e_s = \vec e_x \cos \alpha + \vec e_{ys} \sin \alpha
$ and  a photon in the idler beam (the optical beam the wave vector
$\vec k_i$) with a polarization in the direction $\vec e_i = \vec
e_x \cos \beta + \vec e_{yi} \sin \beta $ is proportional to
\begin{equation}
    \langle \psi |
    [ \vec e_i \cdot \vec {\cal E}_i^\dag (\vec r)]
    [\vec e_s \cdot \vec {\cal E}_s^\dag (\vec r)]
    [\vec e_s \cdot \vec {\cal E}_s (\vec r)]
    [ \vec e_i \cdot \vec {\cal E}_i (\vec r)]
    | \psi \rangle .
\end{equation}

By applying relations (\ref{stt5}),(\ref{fld1})and(\ref{fld2}) we
obtain
\begin{eqnarray}
    P(\alpha,\beta) &\propto&
    \left|
    \left(\cos \alpha + \frac{k_{yd}\cot \theta }{n_o k_0} \sin \alpha \right)
    \left(\sin \beta + \frac{k_{yd}\cot \theta }{n_o k_0} \cos \beta \right)
    \right.
    \nonumber \\
    &&\left.
    + \left(\cos \beta - \frac{k_{yd}\cot \theta }{n_o k_0} \sin \beta \right)
    \left(\sin \alpha - \frac{k_{yd}\cot \theta }{n_o k_0} \cos \alpha \right)
    \right|^2
     \nonumber \\
    &=&\left[1+ \left(\frac{k_{yd}\cot \theta }{n_o k_0} \right)^2 \right]
    (\cos \alpha \sin \beta + \cos \beta \sin \alpha)^2
    \nonumber \\
    &\propto& \sin^2(\alpha + \beta).
\end{eqnarray}

One may observe that by inserting a suitable wave plate into the
signal (or idler) beam to swap the x- and y-component of the optical
field, or to introduce a phase difference equal to $\pi$ between
these two components, correlations like $\sin^2(\alpha - \beta)$,
$\cos^2(\alpha + \beta)$, $\cos^2(\alpha - \beta)$ can also be
obtained. All these correlations have been observed in experiments
\cite{kwt}, but they are not real evidences for quantum
non-locality, because those two photons generated from SPDC are not
spatially separated, and both of them can be found in the same
signal or idler beam at the same time. The probability $P_s$ of
finding both photons in the signal can be calculated by using
relations (\ref{stt5}) and (\ref{fld1}), we have
\begin{eqnarray}
    P_s(\alpha,\alpha^\prime) &\propto&
    \langle \psi |
    [ \vec e_s^{\,\prime} \cdot \vec {\cal E}_s^\dag (\vec r)]
    [\vec e_s \cdot \vec {\cal E}_s^\dag (\vec r)]
    [\vec e_s \cdot \vec {\cal E}_s (\vec r)]
    [ \vec e_s^{\,\prime} \cdot \vec {\cal E}_s (\vec r)] | \psi \rangle
    \nonumber \\
    &\propto&
    \left|
    \left(\cos \alpha + \frac{k_{yd}\cot \theta }{n_o k_0} \sin \alpha \right)
    \left(\sin \alpha^\prime - \frac{k_{yd}\cot \theta }{n_o k_0}
    \cos \alpha^\prime \right)
    \right.
    \nonumber \\
    &&\left.
    + \left(\cos \alpha^\prime + \frac{k_{yd}\cot \theta }{n_o k_0}
    \sin \alpha^\prime \right)
    \left(\sin \alpha - \frac{k_{yd}\cot \theta }{n_o k_0} \cos \alpha \right)
    \right|^2
     \nonumber \\
    &\propto & \sin^2 (\alpha + \alpha^\prime - \gamma) ,
\end{eqnarray}
where $\vec e_s = \vec e_x \cos \alpha + \vec e_{ys} \sin \alpha $
and $\vec e_s^{\,\prime} = \vec e_x \cos \alpha^\prime + \vec e_{ys}
\sin \alpha^\prime $ are two vectors of unity in the polarization
directions of two photons, and
\begin{equation}
    \gamma = \tan ^{-1}\frac{2 k_{yd}}{\sqrt{n_o^2 k_0^2 \tan^2 \theta
    - k^2_{yd}}}.
\end{equation}

\section{Conclusion}

We presented a quantum theory for the quantum state of photon pairs
generated from SPDC, in which the effect of the finite size of
optical crystals is explicitly taken into consideration. Explicit
expressions for the quantum state and optical fields of these photon
pairs are obtained, and we found that these photon pairs are not in
entangled quantum states. We analyzed the correlation between the
polarizations of the photon in signal beam and the photon in the
idler beam, and we showed that this correlation is not an evidence
for quantum non-locality as the two photons generated from SPDC are
not spatially separated. Results on quantum state of photon pairs
generated from SPDC obtained in this paper can also be applied to
explain other apparently non-local phenomena, such as ``ghost''
interference and diffraction \cite{ghst}, four photon entanglement
\cite{4ent, 4ghz} and De Broglie wavelength \cite{4dw}. We will
present those works in separate papers.

\end{document}